\makeatletter
\declare@file@substitution{revtex4-1.cls}{revtex4-2.cls}
\makeatother

\documentclass[twocolumn, times]{aastex631} 
\usepackage{natbib}
\usepackage{amsmath}
\usepackage{amssymb}
\usepackage{subfigure}
\usepackage{url}
\usepackage{longtable}

\renewcommand{\vec}[1]{ {\mathbf #1} }

\newcommand{\Eq}{{Equation}}

\newcommand{\Fig}{{Figure}}
\newcommand{\Figs}{{Figures}}

\graphicspath{figs/}

\shorttitle{Simulation of AR~12158}
\shortauthors{Jiang et al.}

\newcommand{\PaperI}{\href{https://arxiv.org/abs/2308.06977}{Paper I}}

\begin{document}

\title{Magnetic Reconnection as the Key Mechanism in Sunspot Rotation Leading to Solar Eruption}

\author[0000-0002-0786-7307]{Chaowei Jiang} \affiliation{Shenzhen Key
  Laboratory of Numerical Prediction for Space Storm, Institute of
  Space Science and Applied Technology, Harbin Institute of
  Technology, Shenzhen 518055, China} \affiliation{Key Laboratory of
  Solar Activity and Space Weather, National Space Science Center,
  Chinese Academy of Sciences, Beijing 100190, China}

\author{Xueshang Feng} \affiliation{Shenzhen Key Laboratory of
  Numerical Prediction for Space Storm, Institute of Space Science and
  Applied Technology, Harbin Institute of Technology, Shenzhen 518055,
  China} \affiliation{Key Laboratory of Solar Activity and Space
  Weather, National Space Science Center, Chinese Academy of Sciences,
  Beijing 100190, China}

\author{Xinkai Bian} \affiliation{Shenzhen Key Laboratory of Numerical
  Prediction for Space Storm, Institute of Space Science and Applied
  Technology, Harbin Institute of Technology, Shenzhen 518055, China}

\author{Peng Zou} \affiliation{Shenzhen Key Laboratory of Numerical
  Prediction for Space Storm, Institute of Space Science and Applied
  Technology, Harbin Institute of Technology, Shenzhen 518055, China}

\author{Aiying Duan} \affiliation{School of Atmospheric Sciences, Sun
  Yat-sen University, Zhuhai 519000, China}

\author{Xiaoli Yan} \affiliation{Yunnan Observatories, Chinese Academy
  of Sciences, Kunming 650216, China}

\author{Qiang Hu} \affiliation{Center for
  Space Plasma and Aeronomic Research, The University of Alabama in
  Huntsville, Huntsville, AL 35899, USA}

\author{Wen He}\affiliation{Center for Space Plasma and Aeronomic
  Research, The University of Alabama in Huntsville, Huntsville, AL
  35899, USA}

\author{Xinyi Wang} \affiliation{Key Laboratory of Solar Activity and
  Space Weather, National Space Science Center, Chinese Academy of
  Sciences, Beijing 100190, China}

\author{Pingbing Zuo} \affiliation{Shenzhen Key Laboratory of
  Numerical Prediction for Space Storm, Institute of Space Science and
  Applied Technology, Harbin Institute of Technology, Shenzhen 518055,
  China} \affiliation{Key Laboratory of Solar Activity and Space
  Weather, National Space Science Center, Chinese Academy of Sciences,
  Beijing 100190, China}

\author{Yi Wang} \affiliation{Shenzhen Key Laboratory of Numerical
  Prediction for Space Storm, Institute of Space Science and Applied
  Technology, Harbin Institute of Technology, Shenzhen 518055, China}
\affiliation{Key Laboratory of Solar Activity and Space Weather,
  National Space Science Center, Chinese Academy of Sciences, Beijing
  100190, China}

\begin{abstract}
  The rotation of sunspots around their umbral center has long been
  considered as an important process in leading to solar eruptions,
  but the underlying mechanism remains unclear.  A prevailing physical
  picture on how sunspot rotation leads to eruption is that, by
  twisting the coronal magnetic field lines from their footpoints, the
  rotation can build up a magnetic flux rope and drive it into some
  kinds of ideal magnetohydrodynamics (MHD) instabilities which initiate eruptions. Here with a
  data-inspired MHD simulation we studied the rotation of a large sunspot in solar active region NOAA 12158 leading to a
  major eruption, and found that it is distinct from prevailing theories based on
  ideal instabilities of twisted flux rope. The simulation suggests that, through
  successive rotation of the sunspot, the coronal magnetic field is
  sheared with a central current sheet created progressively within the sheared arcade before the eruption, but without forming a flux rope.
  Then the eruption is instantly triggered once fast reconnection sets in at the current sheet, while a highly twisted flux rope is created during the eruption. Furthermore, the simulation reveals an intermediate evolution stage between the quasi-static energy-storage phase and
  the impulsive eruption-acceleration phase. This stage may correspond to the slow-rise
  phase in observation and it enhances building
  up of the current sheet.
\end{abstract}


\section{Introduction}
\label{sec:intro}

Magnetic fields play a defining role for solar activities, especially,
solar eruptions such as solar flares and coronal mass ejections
(CMEs). The most visible manifestation of solar magnetic field are
sunspots as seen on the solar surface (namely, the photosphere), which
represent regions where the strongest magnetic field protrudes from
the solar interior into the atmosphere. When rotating along with the
solar surface, sunspots are also commonly observed to be rotating
around their umbral
center~\citep{brownObservationsRotatingSunspots2003,
  zhangInteractionFastRotating2007,
  yanRelationshipRotatingSunspots2008, minRotatingSunspotAR2009,
  jamesNewTriggerMechanism2020}, which has been discovered a century
ago~\citep{evershed.RadialMovementSunspots1910,st.johnRadialMotionSunSpots1913}. Such
rotational motion of sunspots has long been considered as an important
process in association with generation of solar eruptions, because it
is an efficient mechanism for transporting free magnetic energy and
helicity from below the photosphere into the
corona~\citep{stenfloMechanismBuildupFlare1969,
  barnesForceFreeMagneticFieldStructures1972}. Indeed, almost all the
flare-productive solar active regions (ARs) have been reported with
significant sunspot rotations
\citep{yanRelationshipRotatingSunspots2008}, for example, in the
extensively studied ARs including NOAA 10930
\citep{minRotatingSunspotAR2009,zhangInteractionFastRotating2007},
11158
\citep{jiangRAPIDSUNSPOTROTATION2012,vemareddyROLEROTATINGSUNSPOTS2012},
11429 \citep{zhengSunspotsRotationMagnetic2017}, 12158
\citep{biObservationReversalRotation2016,
  vemareddySUNSPOTROTATIONDRIVER2016}, and 12673
\citep{yanSimultaneousObservationFlux2018,yanSuccessiveXclassFlares2018},
etc.

A widely-believed physical picture on how sunspot rotation leads to
eruption is that, by twisting the coronal magnetic field lines at
their footpoints, the rotation can build up a magnetic flux rope (MFR,
which is a bundle of magnetic flux possessing a significant amount of
twist, typically characterized by a field-line twist number above
unity \citep{chenPhysicsEruptingSolar2017,
  chengOriginStructuresSolar2017, liuMagneticFluxRopes2020}), and
drive it into some kinds of ideal magnetohydrodynamics (MHD)
instabilities \citep[for example, the kink instability and torus instability,][]{kliemTorusInstability2006,
  torokConfinedEjectiveEruptions2005, fanOnsetCoronalMass2007,
  aulanierFORMATIONTORUSUNSTABLEFLUX2010} or catastrophic loss of
equilibrium \citep{forbesCatastropheMechanismCoronal1991,
  linEffectsReconnectionCoronal2000,
  kliemCATASTROPHEINSTABILITYERUPTION2014}. For example, in the
comprehensive review book ``New Millennium Solar
Physics''~\citep{aschwandenNewMillenniumSolar2019}, a section is
devoted to the studies of rotating sunspots and it is written that ``The
relationship between rotating sunspots and the triggering of a flare
accompanied by a sigmoid eruption, most likely driven by a kink
instability, is overwhelming.''

It seems evident that the rotational motion of magnetic field line
footpoints (around the rotating center) can increase magnetic twist
degree of the flux rope (around the rope axis that roots at the
rotating center), until it reaches a critical value for kink
instability \citep{hoodKinkInstabilitySolar1979,
  mikicDynamicalEvolutionTwisted1990,
  galsgaardHeatingActivitySolar1997}. Or, the flux rope will expand
upward owing to the increase of magnetic pressure as driven by the
rotation \citep{aulanierEquilibriumObservationalProperties2005}, and
reaches a critical height at which the torus instability
\citep{kliemTorusInstability2006} (or equivalently, the catastrophic
loss of equilibrium \citep{kliemCATASTROPHEINSTABILITYERUPTION2014})
sets in. However, none of these scenarios has been proven in MHD
simulations that start from a magnetic arcade and are driven by solely
line-tied surface rotation motion without other requirements (for
example, flux cancellation
\citep{amariCoronalMassEjection2003a,aulanierFORMATIONTORUSUNSTABLEFLUX2010}). For
example, all previous attempts of such type of simulations
\citep{amariVeryFastOpening1996, torokEvolutionTwistingCoronal2003,
  aulanierEquilibriumObservationalProperties2005} show that the
continuous twisting of the core field in a bipolar potential field can
lead to a strong expansion (or ``fast opening'') of the field, but
such expansion cannot be taken as solar eruption since there is no
impulsive release (increase) of magnetic (kinetic) energy, and the expansion process
can always relax smoothly to an equilibrium if the
driving velocities are suppressed, therefore not associated with an
instability or a loss of equilibrium
\citep{aulanierEquilibriumObservationalProperties2005}.  There is only
one simulation~\citep{torokInitiationCoronalMass2013} designed to show
that sunspot rotation can cause the arcade overlying a pre-existing
flux rope to inflate, thus weakening the confining effect on the flux
rope and letting it to ascend slowly until reaching the torus
instability. But such a simulation does not show how the flux rope
forms, and moreover the rotating sunspot does not energize the key
structure of eruption (i.e., the flux rope, which is not rooted in the rotating sunspot in their simulation), therefore not playing a
direct role in triggering the eruption.

In a previous work~\citep[][hereafter referred to as \PaperI]{Jiang2023arXiv}, with a data-driven MHD simulation, we have modelled the generation of a major eruption (an X-class eruptive flare) in solar AR NOAA~12158, which contains a continuously rotating sunspot. In that simulation, the photospheric velocity as recovered from the time sequence of vector magnetograms are input directly as the driving flow at the bottom boundary of the model. The simulation successfully reproduced a transition from the quasi-static state to an eruptive phase with onset time matching rather well with observation. It further showed that the mechanism leading to eruption is different from the aforementioned ones, but follows the one as recently established for an idealized bipolar magnetic
configuration~\citep{jiangFundamentalMechanismSolar2021,
  bianHomologousCoronalMass2022,
  bianNumericalSimulationFundamental2022}, in which a current sheet is
formed by quasi-statically shearing the bipolar arcade until
fast reconnection at the current sheet triggers an eruption.  


Although the data-driven simulation is more realistic than other types of simulation, it cannot clearly pin down which photospheric motion plays the key role in driving the field to its eruption. That is, it remains unclear what is the role played by the rotation of sunspot in the process of leading to the eruption. To this end, in this paper we performed a data-inspired simulation for which the driving flow at the photosphere is specified as a rigid rotation profile of the sunspot. Such conceptual simulation is not aimed to realistically reproduce the observation but can provide important clues on which effect (here, the sunspot rotation) is important in leading to eruption. We have also proposed a new way to estimate the degree of sunspot rotation from observations, which is used for guiding the data-inspired simulation. Essentially the data-inspired simulation confirms the same initiation mechanism as revealed in Paper~I. Moreover, attention is also paid to an intermediate evolution stage that transforms the quasi-static phase to the impulsive eruption-acceleration phase. This stage may correspond to the slow-rise phase in observation and it enhances building up of the current sheet.
In the following, we first give an observational
analysis of the studied event in Section~\ref{obs}, then briefly describe the
numerical modeling  in Section~\ref{model} and show the
simulation results in Section~\ref{res}. Finally we conclude and give
discussions in Section~\ref{con}.

\section{Observations}\label{obs}

\begin{figure*}[htbp]
  \centering \includegraphics[width=\textwidth]{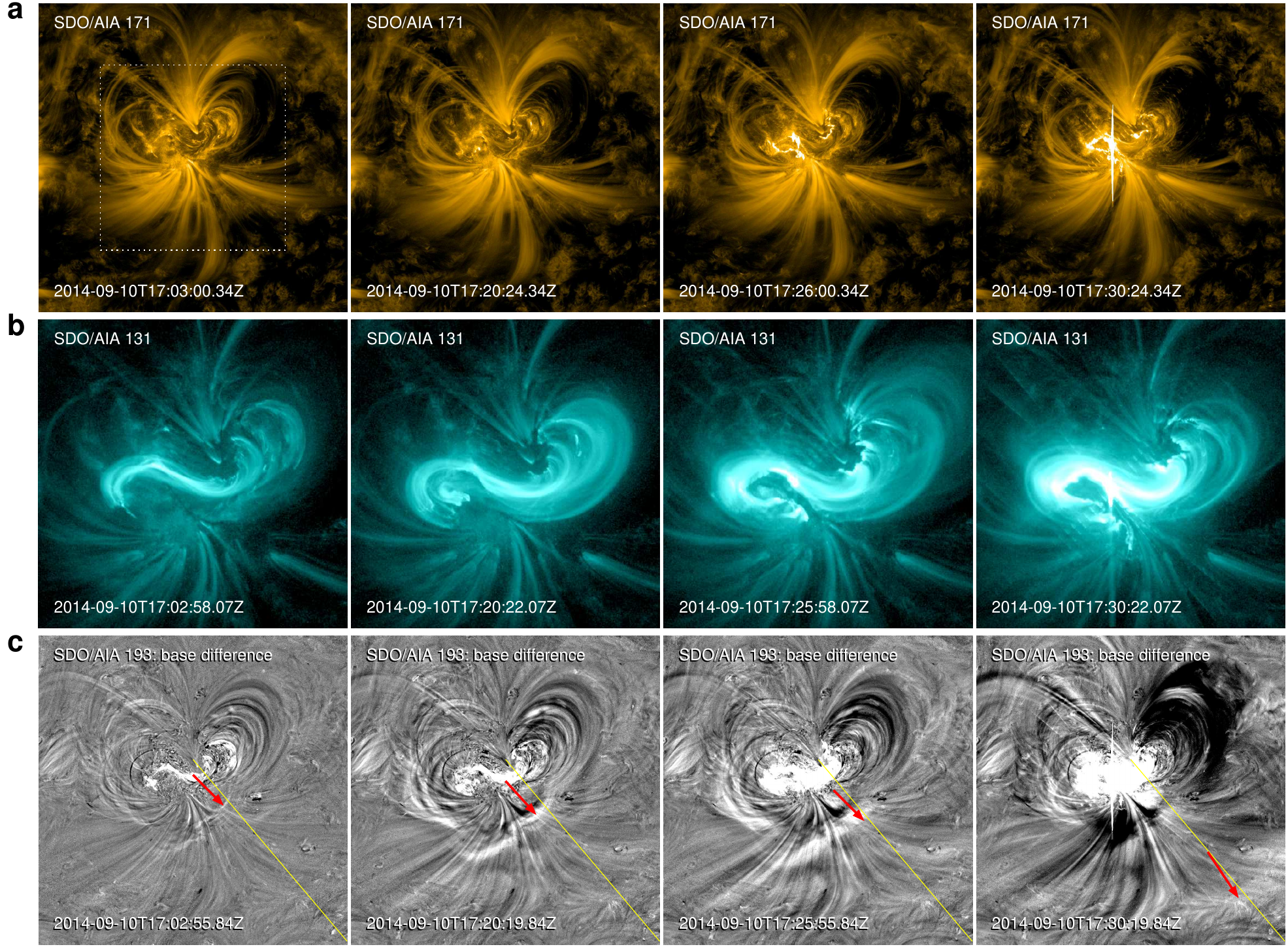}
  \caption{\textbf{EUV imaging of the X1.6 flare taken by SDO/AIA.}
    \textbf{a}, Images in 171~{\AA} channel. \textbf{b}, 131~{\AA}
    channel, with a smaller field of view (as denoted by the dashed
    box in \textbf{a}) to show the core region of the flare site.
    \textbf{c}, Base difference images of the 193~{\AA} channel. The
    yellow line denotes the slit for which the time-distance map in
    \Fig~3b is shown. The red arrows show the moving of an erupting
    loop-like structure that experienced first a slow rise well before
    (for around 20~min) the flare onset time and then an impulsive
    acceleration in the first few minutes during the flare, with more
    details shown in \Fig~\ref{eruption_stack}. The animation attached shows the flare eruption observed by SDO/AIA in different EUV wavelengths
and the associated soft X-ray light curve recorded by GOES. In particular, the first two top panels and the bottom middle panel correspond to the three panels in this figure, respectively. The top right panel shows  soft X-ray light curve. The bottom left panel shows the AIA observation in 193~{\AA}. The bottom right panel shows the time-distance stack plot for the slit as shown in the bottom middle panel.}
   \label{eruption_aia}
\end{figure*}

When first appearing on the solar disk on 5 September 2014, AR
NOAA~12158 was already in its decaying phase, while its leading sunspot
witnessed with continuous counter-clockwise rotation from
6--11 September 2014 during its passage on the solar disk. As a result, the core field of the AR forms a prominent sigmoid in the corona (\Fig~\ref{eruption_aia}). The AR produced a major flare of Geostationary-Operational-Environmental-Satellite (GOES) X1.6 at around 17:00~UT on 10 September.
It is a global eruption of the AR resulting in a halo
CME~\citep{vemareddySUNSPOTROTATIONDRIVER2016}.

\begin{figure*}[htbp]
   \centering
   \includegraphics[width=0.6\textwidth]{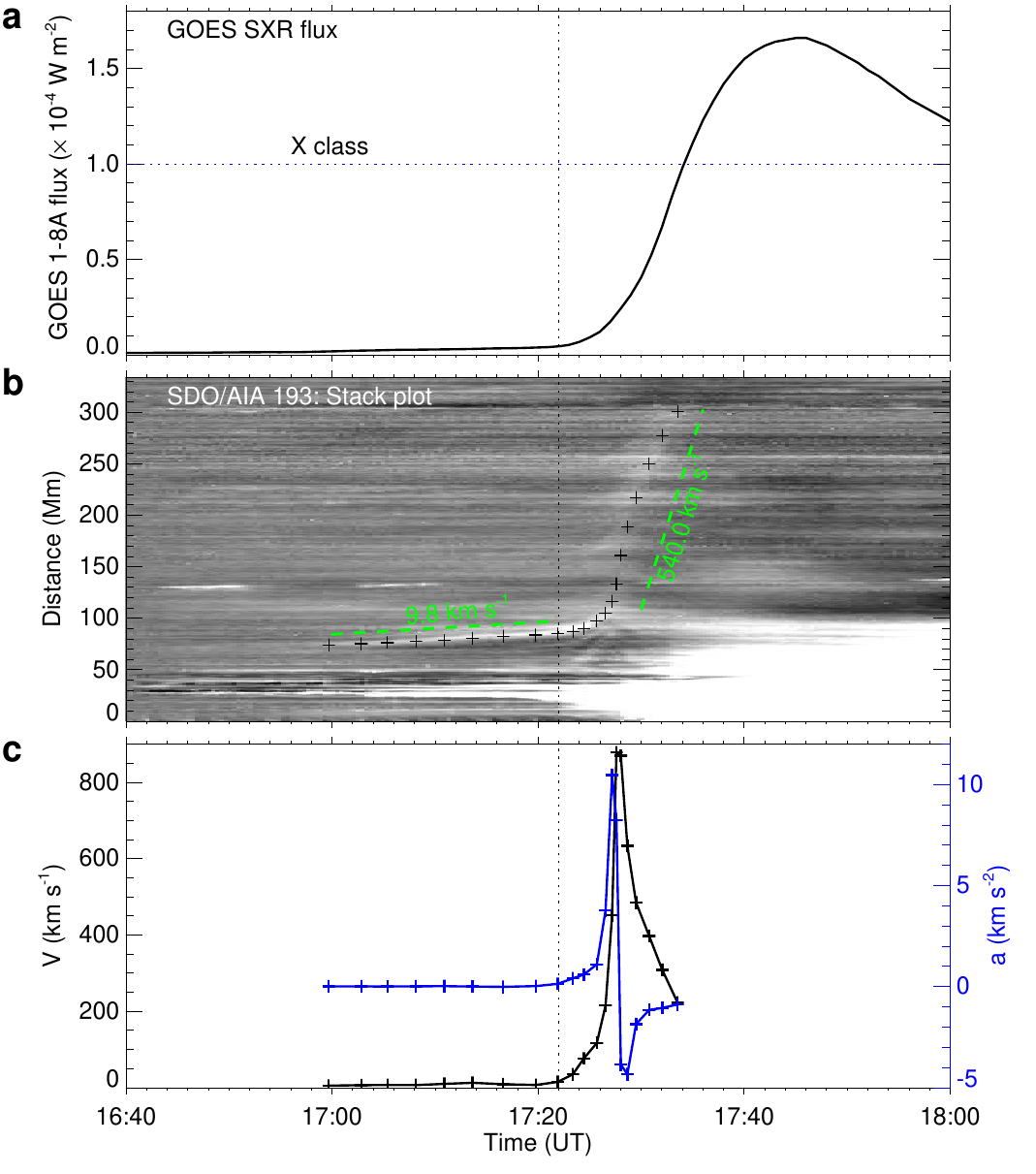}
   \caption{\textbf{Slow rise and impulsive acceleration of the
       erupting coronal loops.} \textbf{a}, GOES soft X-ray flux
     observed in the time period around the X1.6 flare.  The vertical
     dashed line denotes the start of the fast rise of the flux. The
     horizontal dashed line denotes the level of flux that is defined
     for flare reaching X class. \textbf{b}, Time-distance stack plot
     for the slit as shown in \Fig~\ref{eruption_aia}c. The plus signs overlaid are
     sampled to get the data of time and distance of the erupting
     loop-like structure, which shows first a slow rise and then an
     impulsive acceleration. The average speeds of the slow rise phase
     and the eruption phase are denoted by the green dashed
     lines. \textbf{c}, Evaluation of velocity and acceleration of the
     erupting structure as shown in \textbf{b} based on the sampled
     data points.}
   \label{eruption_stack}
\end{figure*}

\subsection{Observation of the eruption}

\Fig~\ref{eruption_aia} (and its animation) shows the eruption process
imaged in three AIA channels. With about 20~min before the flare
onset, a precursor is observed; in the hot channel of AIA~131~{\AA}, a
few loops in the core of the sigmoid became more and more prominent,
while the transverse width of this brightening structure became
progressively thinner, which possibly hints that a pre-flare current
sheet was forming gradually there. Meanwhile, a set of coronal loops,
likely overlying the middle part of the sigmoid, expanded outward
slowly, which looks rather faint but still detectable in different AIA
channels and can be clearly seen in the base difference images of
AIA~193~{\AA} (\Fig~\ref{eruption_aia}c). The speed in the direction
from the flare core to the southwest is estimated to be around
10~km~s$^{-1}$ (\Fig~\ref{eruption_stack}), which is at least an order
of magnitude larger than a quasi-static evolutionary speed that is
driven by the photospheric motions (note that the actual expansion
speed of these loops is underestimated due to the projection effect).
At 17:21~UT, the soft X-ray flux increased impulsively
(\Fig~\ref{eruption_stack}a), which indicates the onset of fast
reconnection. Instantly, the slow-expanding loops were accelerated
impulsively within about 3 min, reaching a speed of
$\sim 800$~km~s$^{-1}$ at 17:26~UT (\Fig~\ref{eruption_stack}b and
c). The peak value of acceleration reaches above
$\sim 10$~km~s$^{-2}$, which is on the same order of magnitude of
acceleration as observed in the most impulsive eruption events. Such
strong acceleration is not likely driven by an ideal MHD instability
of MFR, according to a study by
~\citet{vrsnakProcessesMechanismsGoverning2008}, who used a analytic
model of flux rope to study the acceleration of eruption as driven by
ideal evolution and resistive evolution. They concluded that ``The
purely ideal MHD process cannot account for highest observed
accelerations which can attain values up to $10$~km~s$^{-2}$. Such
accelerations can be achieved if the process of reconnection beneath
the erupting flux-rope is included into the model.''~\citep[see
also][]{greenOriginEarlyEvolution2018}.

With start of the impulsive phase of the flare, the flare ribbons and
post-flare loops were seen with slipping-like motions, which are
interpreted to be a manifestation of 3D slipping reconnection that
forms an MFR during flare~\citep{liQUASIPERIODICSLIPPINGMAGNETIC2015,
  dudikSLIPPINGMAGNETICRECONNECTION2016,gouCompleteReplacementMagnetic2023},
and a twin set of coronal dimmings were observed in both AIA~171 and
193~{\AA} channels along with the eruption, which should map the feet
of the erupting MFR. Although some lowing-lying filaments were
observed at the flare
site~\citep{dudikSLIPPINGMAGNETICRECONNECTION2016}, they did not erupt
with the flare. Previous studies of the same event using coronal
nonlinear force-free field extrapolation also suggest that only a
sheared arcade rather than a well-defined MFR existed prior to this
eruption~\citep{duanComparisonTwoCoronal2017,
  shenPrecursorPhaseXclass2022}.

\begin{figure*}[htbp]
   \centering
   \includegraphics[width=\textwidth]{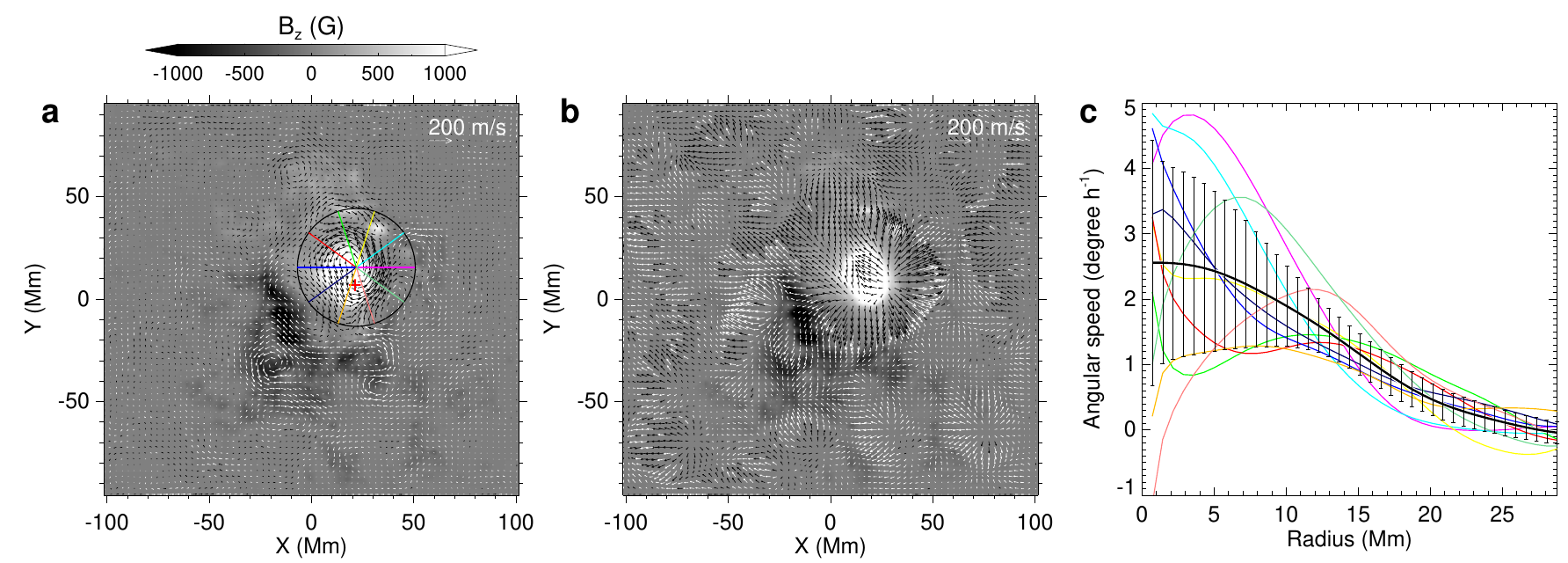}
   \caption{\textbf{Estimation of the rotation speed of the sunspot.}
     \textbf{a}, Vectors show the divergence-free component $\vec v_1$
     of the surface flow averaged for the three-day data covering 2014
     September 8-10. The background image shows the magnetic flux
     density $B_z$, also averaged for the three-day HMI data. The
     circle denotes the rotating sunspot, and the centre of the circle
     is the rotating center. The radial lines are sampled to estimate
     the profile of angular speed of different radial distances from
     the circle center. Here 10 lines are shown and in the actual
     computation we used 100 lines with azimuthal angle distributed
     evenly from $0$ to $2\pi$. The red plus symbol marks the point
     with the largest $B_z$ in the sunspot. It can be regarded as the
     center of the sunspot and it is close to the rotation center.
     \textbf{b}, Same as \textbf{a} but for the rest of the surface
     flow, i.e., the curl-free component $\vec v_2$.  \textbf{c},
     Profile of angular speed at different radial distance from the
     rotation center. The thin coloured lines denote the values along
     the corresponding radial lines (by the same colours) as shown in
     \textbf{a}. The thick line (with error bars) denotes the average value (and standard deviation) of angular speed for all the azimuthal angles from $0$ to $2\pi$.}
   \label{fig_rot_sup}
\end{figure*}

\subsection{Estimation of sunspot rotation}
To guide the simulation, we first estimate the degree of the sunspot
rotation in the three days of 8--10 September 2014, as our simulation will be started with initial conditions representing the MHD equilibrium at the time of 00:00~UT on 8 September 2014. To estimate the
degree of its rotation with respect to the sunspot center, we
calculate the surface velocity at the photosphere based on a time
sequence of vector magnetograms, and further extract the velocity
components that are directly associated with the rotational motion
from the surface flow. Then, by integration of the rotational speed
with time, it provides an alternative way of estimating the rotational
degree to the traditional method based on direct inspection of the
white-light continuum
images~\citep{brownSemiAutomaticMethodMeasure2021,
  vemareddySUNSPOTROTATIONDRIVER2016}. 

The surface velocity at the
photosphere is derived using the DAVE4VM code~\citep{schuckTrackingVectorMagnetograms2008}, which is also used in {\PaperI} where more details can be found.
Based on the photospheric surface velocity as derived, the rotational
speed is estimated in the following way. We first extract the velocity
components relevant only to the rotational motion from the surface
horizontal velocity $\vec v$ by decomposing it as
\begin{equation}
  \label{eq:v}
  \vec v = \vec v_{1} + \vec v_{2} = \nabla \times \vec p + \nabla q,
\end{equation}
where $\vec p$ is a arbitrary vector and $q$ a arbitrary scalar. The curl-free part $\vec v_{2}=\nabla q$ is obtained by solving
a Poisson equation $\nabla^2 q = \nabla \cdot \vec v$, and then the
divergence-free part is
$\vec v_{1}=\nabla \times \vec p = \vec v - \vec v_2$. The rotational
flow of the sunspot is contained only in the divergence-free field
$\vec v_1$. Then we estimated the time average of the rotational flow
of three days by simply averaging the field $\vec v_{1}$ in each pixel
using the three-day data of 8--10 September 2014.
\Fig~\ref{fig_rot_sup}a gives the distribution of the averaged
$\vec v_{1}$, and \Fig~\ref{fig_rot_sup}b shows the averaged
$\vec v_2$. As can be seen, the rotating center is close to the
location with the maximum magnetic flux density in the sunspot (also
on time average).  The curl-free part $\vec v_2$ shows clearly a
diverging flow from approximately the center of the sunspot. With this
averaged flow field, we further calculated the rotational rate, i.e.,
the angular speed of the sunspot with respect to the rotating
center. As shown in \Fig~\ref{fig_rot_sup}c, we sampled 10 radial
lines with different azimuthal angle $\phi$ (evenly distributed from
$0$ to $2\pi$) from the center of rotation, and got the rotational
velocity $v_{\phi}(r, \phi)$ on these radial lines. The average
angular speed is given by
\begin{equation}
  \label{eq:omega}
  \omega(r) = \frac{1}{2\pi}\int \frac{v_{\phi}(r, \phi)}{r} d\phi.
\end{equation}

The angular speed as averaged over three days of 8--10 September 2014
shows that the rotation is the fastest near the center of the sunspot
umbra, reaching around $2.5^{\circ}$~h$^{-1}$, and it decreases
gradually to zero outside of the penumbra at a radial distance of
$25$~Mm (\Fig~\ref{fig_rot_sup}c). When averaged for the whole
sunspot, the rotation speed is about $1.75^{\circ}$~h$^{-1}$, and thus in the three days the sunspot has
rotated with a total degree of $\sim 130^{\circ}$. Our result of the
total rotation degree for the three days from 8--10 September 2014
agrees well with the value obtained from other independent methods for
the same sunspot~\citep{brownSemiAutomaticMethodMeasure2021,
  vemareddySUNSPOTROTATIONDRIVER2016}.


 
\section{Numerical model}\label{model}
Since a detailed description of the numerical model has been provided in \PaperI, including the controlling MHD equation, the grid setting, the initial and the boundary conditions, we will not repeat them here. Same as \PaperI, our simulation is started from an MHD equilibrium constructed for  00:00~UT on 8 September 2014, a time over 65 hours before onset of the X1.6 flare. The simulation aims to follow a long-period of slow coronal magnetic evolution until the major eruption. The key difference in the current simulation with {\PaperI} is that here the simulation is driven by surface motions specified at the bottom boundary by assuming that the major sunspot rotates like a rigid body (namely with a constant angular speed within the sunspot, which is the simplest case) while all other regions are fixed. Therefore, as compared with the data-driven simulation in \PaperI, such a data-inspired simulation can emphasize the effect of sunspot rotation in leading to the eruption.


Specifically, the sunspot is rotated rigidly with respect to its surrounding
field by applying a velocity profile at the bottom boundary defined as
\begin{equation}
  \label{rigidrotv}
  v_x  = -\omega (y-y_c),\ \
  v_y  =  \omega (x-x_c)
\end{equation}
where the rotating center $(x_c, y_c)$ is defined to be the point in
the sunspot with the largest $B_z$, and the angular speed $\omega$ is
a constant (positive, thus rotating counter-clockwise) within a radius
of $10$~arcsec from the rotating center, and then decreases linearly
to zero at a radius of $35$~arcsec. The value of $\omega$ is chosen
such that the maximum value of the surface speed is
$10$~km~s$^{-1}$. 
On the bottom boundary, we
solved the magnetic induction equation to update all the three components
of magnetic field with the flow field prescribed by those defined in
\Eq~(\ref{rigidrotv}), while the plasma density and temperature are
simply fixed.

\begin{figure*}[htbp]
   \centering
   \includegraphics[width=\textwidth]{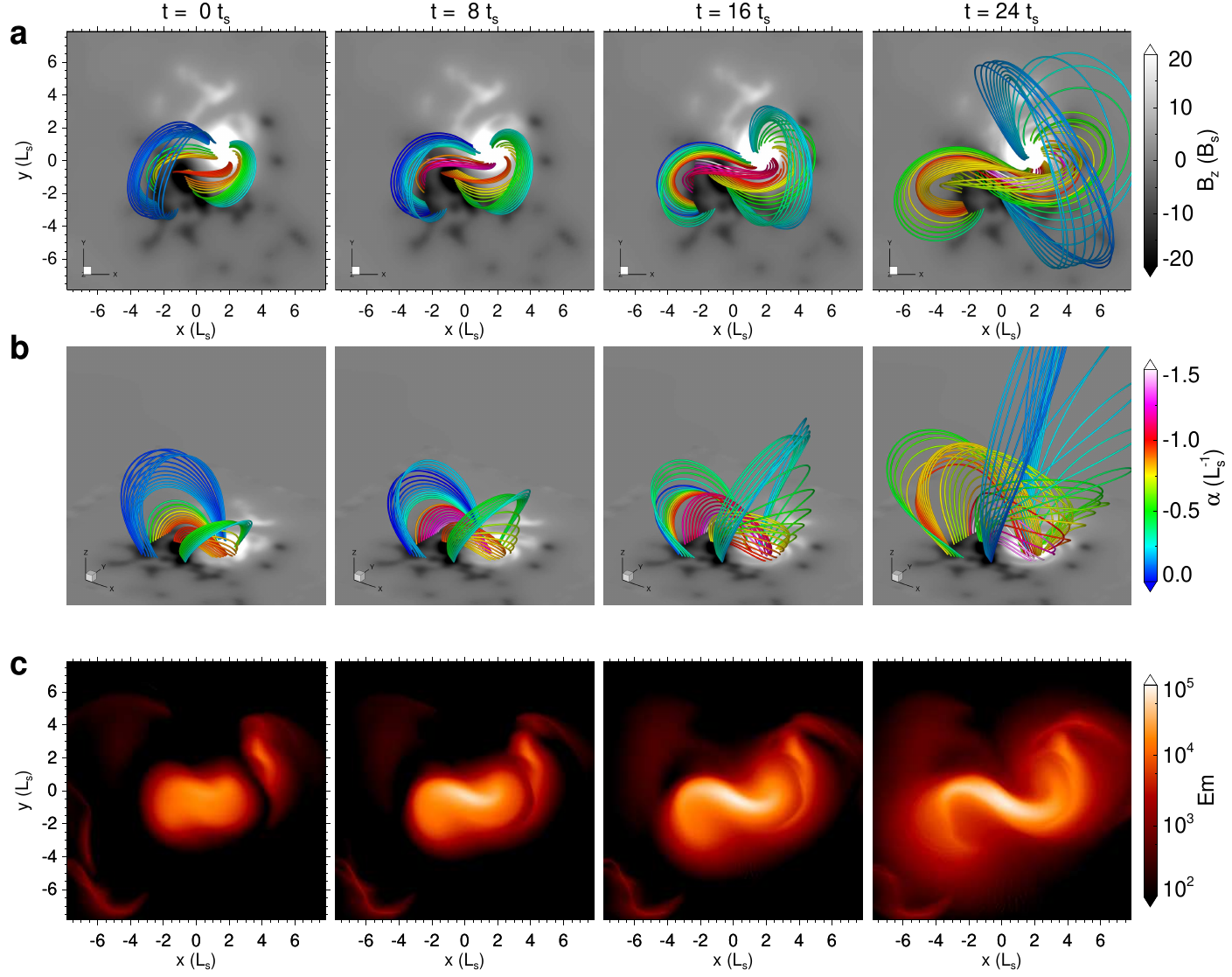}
   \caption{\textbf{Pre-eruption evolution of magnetic field and
       current density in the simulation.} \textbf{a}, Top view of sampled magnetic field
     lines. The coloured thick curves represent magnetic field lines,
     and the colours denote the value of the nonlinear force-free
     factor defined as $\alpha = \vec J\cdot \vec B/B^2$ where
     $\vec J$ is the current density and $\vec B$ is the magnetic
     field. The background shows the magnetic flux distribution on the
     bottom surface. Note that at different times, the field lines are
     traced from the same set of footpoints from the negative
     polarities, since they are fixed without surface motion during
     the simulation with the rigid rotation. $L_s = 11.52$~Mm is the
     length unit, $t_s = 105$~s the time unit, and $B_s = 1.86$~G the
     magnetic field strength unit.  \textbf{b}, 3D prospective view of
     the same field lines shown in \textbf{a}. \textbf{c}, Synthetic
     images of coronal emission from current density. The attached animation shows evolution of magnetic field lines (left), current density (middle),
and magnetic squashing factor (right) in the simulation (for the core region). The left two panels and
correspond to the animation versions of \Fig~\ref{fig3}\textbf{a} and \textbf{b}, respectively, and the middle top panel corresponds to
\Fig~\ref{fig3}\textbf{c}. The bottom middle and right panels correspond to \Fig~\ref{fig3_add}\textbf{a} and \textbf{b}, respectively.
The top right panel corresponds to \Fig~\ref{fig4}\textbf{c}.}
   \label{fig3}
\end{figure*}

\begin{figure*}[htbp]
   \centering
   \includegraphics[width=\textwidth]{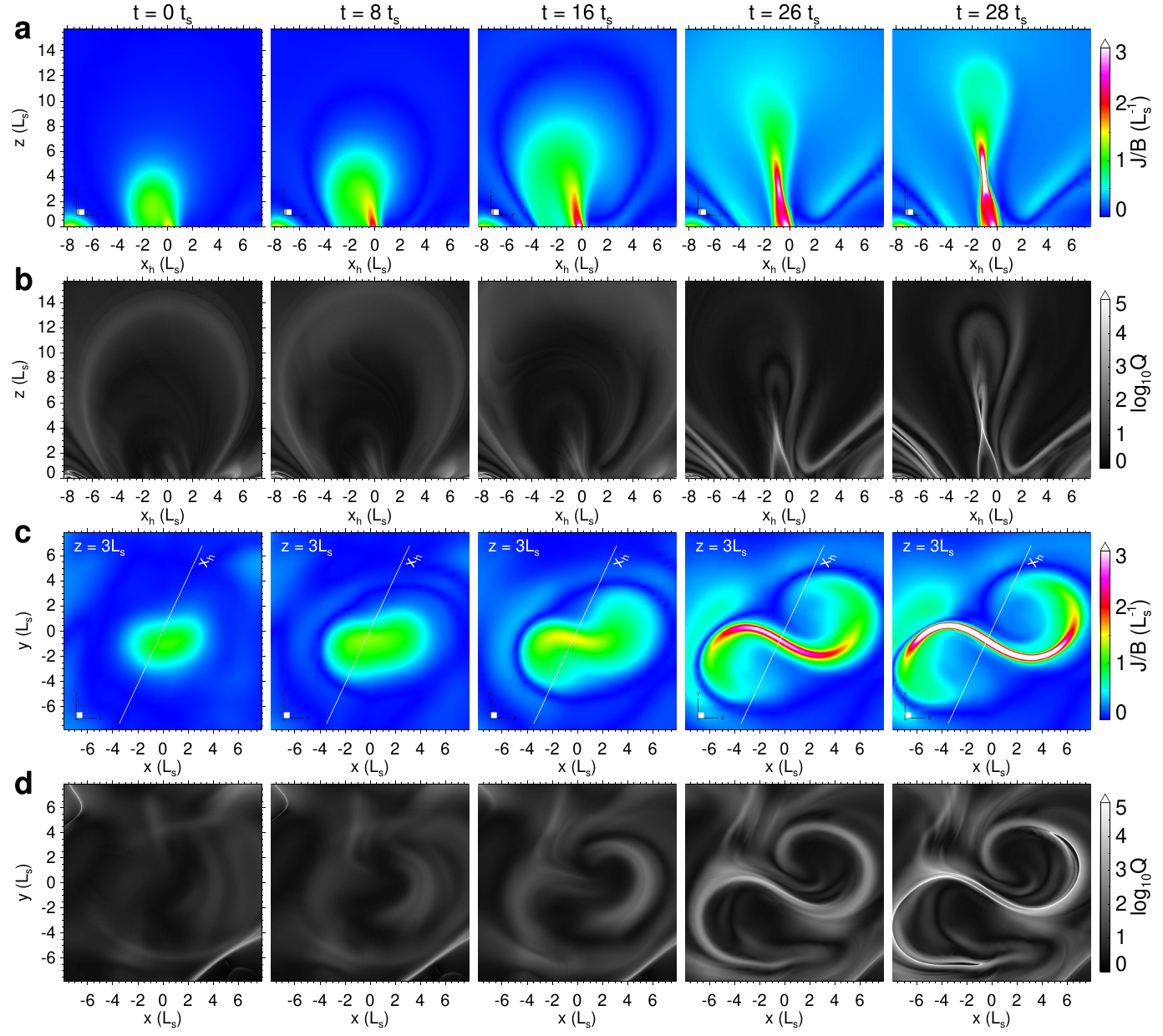}
   \caption{\textbf{Formation of the current sheet in the pre-eruption
       evolution phase of sunspot rigid-rotation simulation.}
     \textbf{a}, Vertical cross-section of the normalized current
     density, namely, $J/B$. \textbf{b}, Distribution of magnetic
     squashing degree (i.e., $Q$ factor) on the same slice of
     \textbf{a}. \textbf{c}, Horizontal cross-section of the
     normalized current density at a fixed height of $z=3L_s$. The
     projected location of the vertical cross-section in \textbf{a}
     and \textbf{b} is denoted by the black line in \textbf{c}, which
     crosses through the current sheet perpendicularly at the point
     with the highest current density. \textbf{d}, Distribution of $Q$
     factor on the same slice of
     \textbf{d}. 
   }
   \label{fig3_add}
\end{figure*}

\begin{figure*}[htbp]
   \centering
   \includegraphics[width=\textwidth]{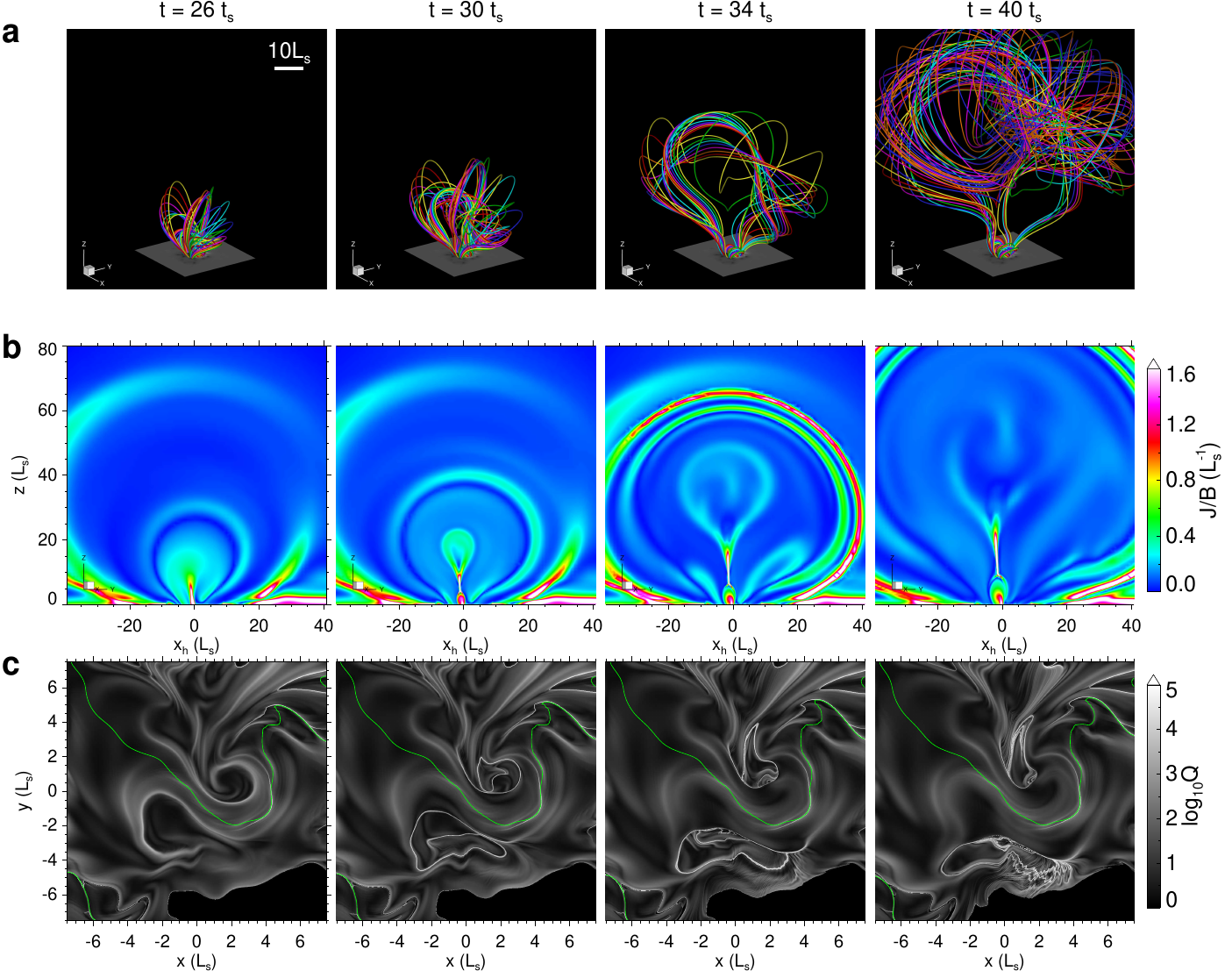}
   \caption{\textbf{Evolution of the erupting field in the sunspot
       rigid-rotation simulation}. \textbf{a}, Magnetic field lines
     are shown by the thick coloured lines, and the colours are used
     for a better visualization of the different lines. The bottom
     surface is shown with the distribution of magnetic flux. The
     field lines at different times are traced from the same set of
     footpoints from the negative polarities, since they are fixed
     without surface motion during the simulation with rigid
     rotation. \textbf{b}, Vertical central cross-section of the
     normalized current density.  Location of the cross section is the
     same one as shown in \Fig~\ref{fig3_add}a. Note that there are
     two thin current layers on top of the erupting MFR (seen at
     $t=34~t_s$), and the shock refers to the outermost one. This
     layer is not related to the topology interface or QSL of the
     field lines. The current layers immediately below the shock is a
     QSL and it is formed well before the shock forms. \textbf{c},
     Magnetic squashing factor at the bottom surface. The green curves
     represent the magnetic polarity inversion line. An animation is attached for
     \textbf{a} and \textbf{b} of this figure.}
   \label{fig4}
\end{figure*}

\begin{figure*}
   \centering
   \includegraphics[width=0.5\textwidth]{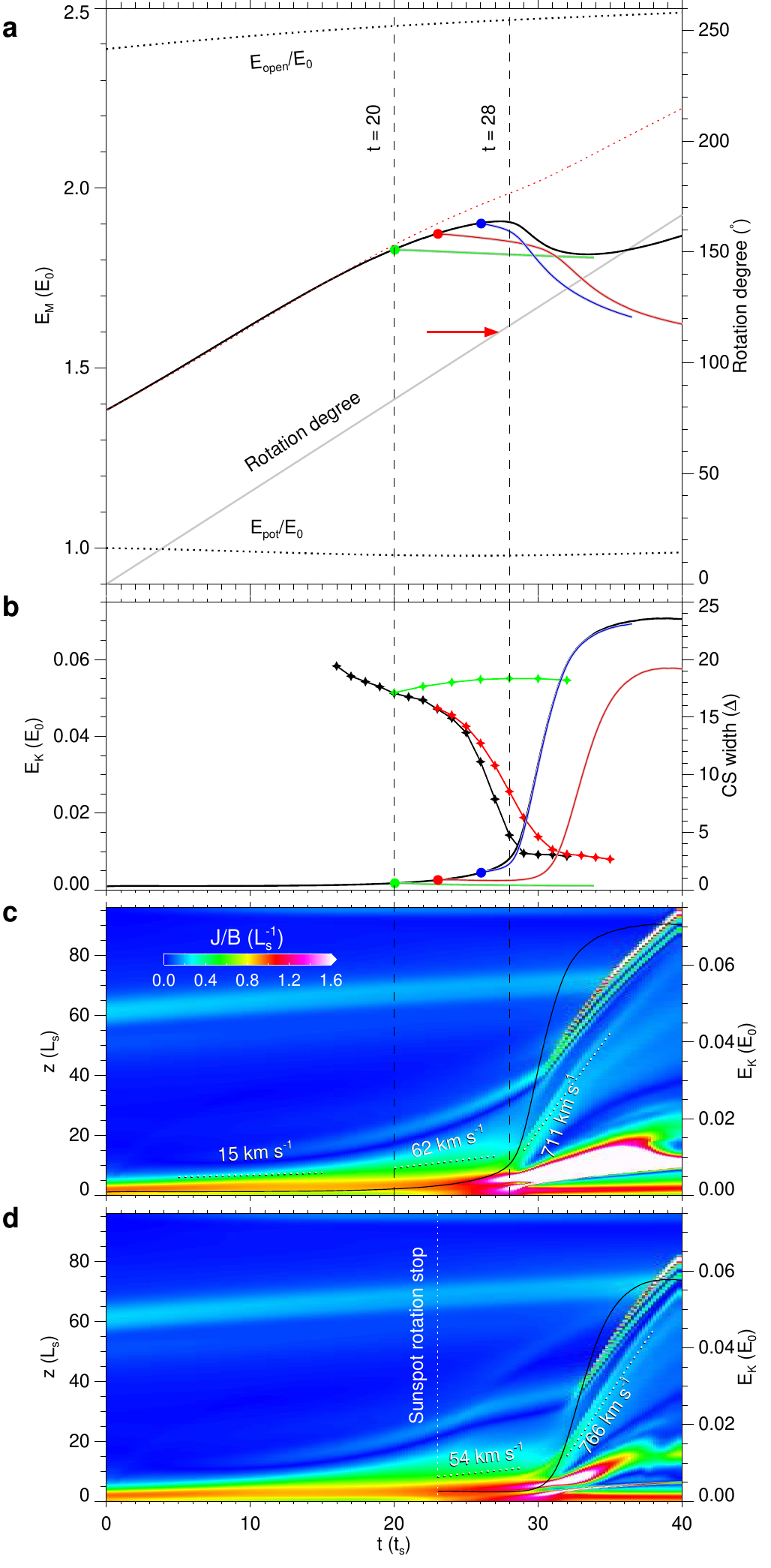}
   \caption{\textbf{Temporal evolution of different parameters in the
       simulation with rigid rotation of sunspot.}  \textbf{a},
     Magnetic energies are shown with the left $y$-axis. The black
     line shows result with continuous rotation. The red dashed line
     shows the energy injected into the volume from the bottom
     boundary through the surface rotational flow, that is, a time
     integration of total Poynting flux at the bottom surface. The
     green, red, and blue lines show results of runs with the rotation
     switched off at $t=20~t_s$, $23~t_s$, and $26~t_s$,
     respectively. All the energies are normalized by the initial
     potential field energy $E_0$, which is $3.04\times 10^{30}$~erg
     in the simulation, and is $1.22\times 10^{33}$~erg if scaled to
     the realistic value. The two black dashed lines show evolution of
     the energies of the open field (the upper one) and the potential
     field (the lower one). The gray line shows the rotational degree
     with the right $y$-axis. The red arrow denotes the total rotation
     degree from the initial time to flare onset as derived from
     observation.  \textbf{b}, Kinetic energies (solid lines) with the
     left $y$-axis, and thickness of the current layer (solid lines
     with stars) with the right $y$-axis. Same as in \textbf{a}, the
     line colored in black, green, red, and blue show results of the
     continuous rotation run, and runs with the sunspot rotation
     stopped at $t=20~t_s$, $23~t_s$, and $26~t_s$, respectively.
     \textbf{c}, A time stack map of the $J/B$ distribution around
     $x, y = 0$ for the continuous rotation run, which is used to show
     the evolution speed of the erupting structure. \textbf{d}, The
     same time stack map of $J/B$ distribution as \textbf{c} but for
     the simulation run with sunspot rotation stopped at
     $t=23~t_s$. The typical speed of the structures are denoted by
     the dashed lines. The animation attached for this figure has the same format
     as the animation of \Fig~\ref{fig4}, but for the simulation with the sunspot rotation
    stopped at $t=23~t_s$.}
   \label{paraevol}
 \end{figure*}

\begin{figure*}[htbp]
   \centering
   \includegraphics[width=0.8\textwidth]{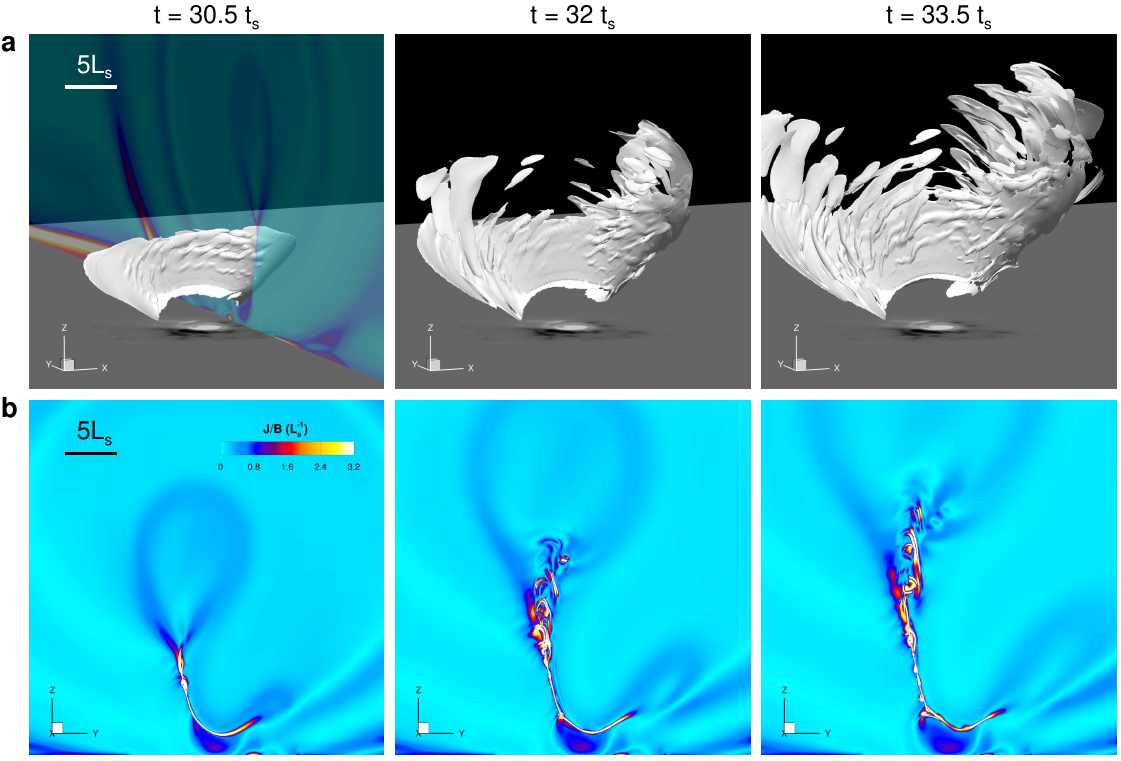}
   \caption{\textbf{A high-resolution simulation of the eruption
       process with the plasmoid instability triggered in the flare
       current sheet.} The grid resolution of the current sheet
     reaches $90$~km, eight times higher than the main
     run. \textbf{a}, The 3D structure of the flare current sheet as
     shown by the iso-surface of $J/B = 3.2 L_s^{-1}$. The bottom
     surface is shown with the distribution of magnetic
     flux. \textbf{b}, A vertical slice of current sheet, for which
     the location is denoted in the first panel of \textbf{a}.}
   \label{turbulent_CS}
\end{figure*}

\section{Simulation results}\label{res}
The simulation demonstrates an evolution from an initially
quasi-static stage to finally a fast eruption. The evolutions of 3D
structure of magnetic field and electric current before the eruption
are shown in \Figs~\ref{fig3} and \ref{fig3_add}, while their
evolutions during eruption are given in \Fig~\ref{fig4}, and the time
profiles of the magnetic and kinetic energies in the whole process,
along with the kinematic behaviour of the erupting features, are
plotted in \Fig~\ref{paraevol}. Below we describe the different stages
and the key processes in the simulated MHD evolution.

\subsection{The pre-eruption stage}
The magnetic configuration at the initial time is a sheared arcade
core enveloped by an overlying, nearly current-free field. As the main
sunspot rotates counterclockwise, the coronal magnetic configuration
expands slowly and the degree of magnetic shear increases, i.e., with
the field lines in the core part becoming more and more aligned with
the bottom PIL as viewed from above (\Fig~\ref{fig3}a and its animation). Progressively,
these field lines, as a whole, form a prominent reverse S shape. 
As shown in \Fig~\ref{fig3}c, a
sigmoid structure is also seen in the synthetic images of coronal
emission from current density, which are obtained by vertical integration of the square of current density averaged along each field line (see also \PaperI). It resembles the observed coronal
sigmoid structure, for example, comparing the last image of
\Fig~\ref{fig3}c with the EUV sigmoid as shown in
\Fig~\ref{eruption_aia}b. In the pre-flare phase from $t=0~t_s$ to
$t=20~t_s$ (where $t_s = 105$~s is the time unit used in the numerical
model), the magnetic energy increases monotonically with a nearly
constant rate (\Fig~\ref{paraevol}a), because the sunspot rotates with
a constant speed. While the magnetic energy rises significantly, the
kinetic energy keeps a small value of around $1\times 10^{-3} E_0$
(where $E_0$ is the potential field energy at the initial time), and
thus most of the injected energy from the bottom boundary through the
surface driving motion (which is indicated by the red dashed line in
\Fig~\ref{paraevol}a) is stored as coronal magnetic energy. Moreover,
since the rotation motion introduces very limited variation to the
magnetic flux distribution at the bottom surface and thus the
corresponding potential-field energy changes only slightly during the
whole evolution process (the black dashed line in \Fig~\ref{paraevol}a), most of the injected energy is stored as free
magnetic energy in the corona. A total amount of free energy of
$\sim 0.5~E_0$ has been stored until the eruption onset at $t=28~t_s$,
and thus the non-potentiality, as measured by the ratio of the total
magnetic energy to the potential field energy, reaches
$1.9$. Meanwhile, the sunspot has rotated by about $115^{\circ}$,
which is almost identical to that derived from observations
($\sim 114^{\circ}$), i.e., an average rotational rate of
$1.75^{\circ}$~h$^{-1}$
multiplied by a time of 65~h before the X1.6 flare. If not interrupted
by the eruption, it seems that with another $100^{\circ}$ of rotation
the non-potentiality can approach an upper limit of approximately
$2.45~E_0$ as determined by the fully open
field~\citep{alyHowMuchEnergy1991,
  sturrockMaximumEnergySemiinfinite1991}. But such ideal evolution is
not possible because a central current sheet unavoidably forms and
triggers reconnection, which results in the eruption.

\subsection{Formation of a current sheet}
A clear signature of current sheet formation can be seen in the
evolution of current density in cross sections of the volume
(\Fig~\ref{fig3_add} and the animation of \Fig~\ref{fig3}). Note that the current density
is normalized by the magnetic field strength (i.e., $J/B$) to
emphasize thin layers with strong current. Initially the current
density is volumetric, and gradually a narrow layer with enhanced
density emerges, becomes progressively thinner. To characterize this evolution of the current layer, we have measured its thickness, which is defined at the location where it is thinnest. 
As can be seen in the variation of the current
layer thickness with time in \Fig~\ref{paraevol}b, the thickness of the current layer decreases all the
way until the onset of the eruption. At the time of
$t = 28~t_s$, the thin current layer extends from the bottom to a
height of $50$~Mm with a thickness of around $3\Delta$ (here
$\Delta = 0.72$~Mm is the finest grid resolution). This is the
critical time point when the current sheet reaches beyond the grid
resolution and the numerical resistivity arises to trigger fast
reconnection in the current sheet, which initiates the eruption. The
current sheet formation is accompanied with the formation of a
quasi-separatrix layer
(QSL~\citep{demoulinThreedimensionalMagneticReconnection1996}) as seen
in the distribution of the magnetic squashing degree (i.e., $Q$
factor, see \Fig~\ref{fig3_add}b and d). The $Q$ factor quantifies the
gradient of magnetic field-line mapping with respect to their
footpoints, and it is helpful for searching topological interface or
QSLs of magnetic flux connections using extremely large values of $Q$
factor ($\sim 10^5$)~\citep{titovTheoryMagneticConnectivity2002,
  liuSTRUCTURESTABILITYEVOLUTION2016}. Initially the $Q$ factor
distributes smoothly with mostly small values. Along with narrowing of
the central current layer, there is an evident increase of $Q$ in the
central thin layer. Immediately prior to the eruption, it has $Q$
reaching $\sim 10^5$ and an extremely small thickness, thus being
identified as QSL, at the same location with the enhanced values of
$J/B$.

\begin{figure*}[htbp]
   \centering
   \includegraphics[width=0.6\textwidth]{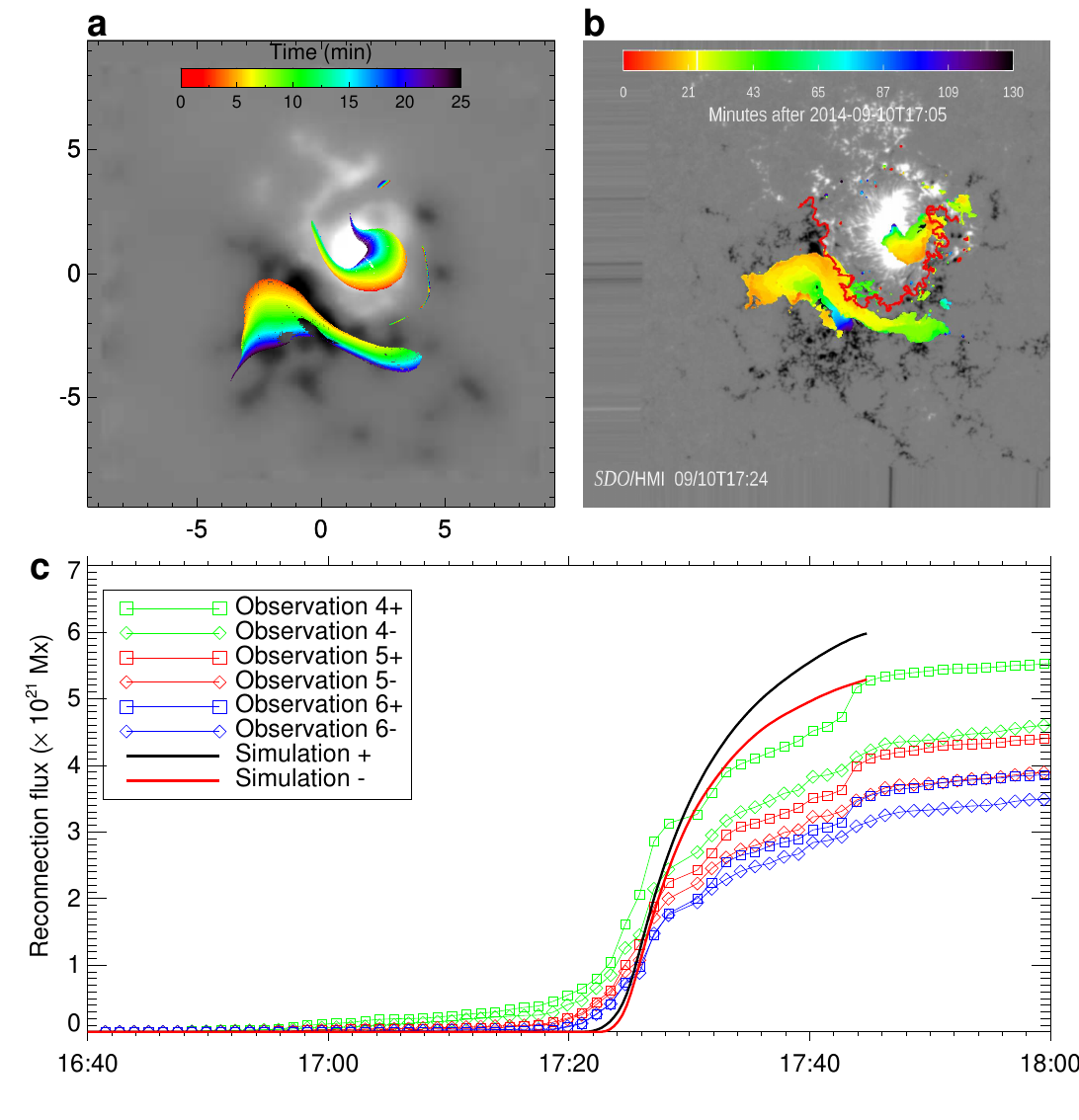}
   \caption{\textbf{Comparison of reconnection fluxes derived from
       simulation and observation.} \textbf{a}, Temporal and spatial
     distribution of simulated ``flare ribbons'', which are the
     footpoint locations of the newly reconnected field lines that
     forms the closed short arcades corresponding to the post-flare
     loops. The color denotes the time of minutes after the simulation
     eruption onset time. The background gray image shows the magnetic
     flux distribution at the bottom surface. \textbf{b}, Format same
     as \textbf{a} but for the observed flare ribbons as imaged in
     SDO/AIA 1600~{\AA} wavelength. \textbf{c}, Evolution of
     reconnection flux, which is the sum of the magnetic flux in the
     area swept by the flare ribbons. For the simulation, the fluxes
     are counted for the positive (the thick black line) and the
     negative (the thick red line) polarities separately. For the
     observed data, three sets of results are obtained by different
     background removal criteria for selection of the flare ribbons,
     which are pixels with the intensity larger than 4 (green line), 5
     (red line), and 6 (blue line) times of the background average
     brightness. The line with boxes (diamonds) represent values
     counted for the positive (negative) polarity.}
   \label{recflux_compare}
\end{figure*}

\subsection{The eruption and reconnection}
With onset of the eruption, the kinetic energy increases impulsively
and reaches finally about $0.07~E_0$, while the total magnetic energy
experiences a fast decrease, even though the boundary driving still
injects magnetic energy into the volume. The total released magnetic
energy amounts to $0.3~E_0$ or $4\times 10^{32}$~erg if scaled to the
realistic value of the magnetic field, which is sufficient to power a
typical X-class flare~\citep{emslieGLOBALENERGETICSTHIRTYEIGHT2012}.
The eruption of the magnetic field creates a large-scale MFR through
the continuous magnetic reconnection in the current sheet
(\Fig~\ref{fig4} and its animation). The existence of such an MFR in
this event has been confirmed by in-situ observation in the
interplanetary space of the CME from this
AR~\citep{kilpuaEstimatingMagneticStructure2021}. Ahead of the MFR,
the eruption drives a fast magnetosonic shock, which is shown by the
thin arc of the current density on the top of the MFR. During the
eruption, the reconnecting current sheet extends in both transverse
size and height, but is kept in the same thickness that is allowed by
the given grid resolution (\Fig~\ref{fig4}b and \Fig~\ref{paraevol}b). When using an
sufficiently high resolution such that the Lundquist number of the
current sheet can reach the order of $10^4 \sim 10^5$, the
reconnecting current sheet would run into plasmoid instability and
could then trigger turbulence, which help to achieve a fast
reconnection
rate~\citep{bhattacharjeeFastReconnectionHighLundquistnumber2009,
  daughtonRoleElectronPhysics2011,
  jiangFundamentalMechanismSolar2021}.  For the current simulation, we
have also carried out an experiment with much higher resolution of
$\Delta = 90$~km, and indeed the plasmoid instability is triggered as
shown in \Fig~\ref{turbulent_CS}. Since the high-resolution run
requires an extremely long computational time, it is performed for
only a short period during the eruption from $t=30.5~t_s$ to
$33.5~t_s$.

Connecting to the bottom of the current sheet is a cusp structure,
below which is the post-flare arcade, i.e., the short field lines
formed after the reconnection. The post-flare arcade and the MFR are
separated, initially partially but later fully, by the QSL that
originates from the current sheet and forms the topological surface of
the MFR~\citep{jiangFormationMagneticFlux2021}. The QSL exhibits two
J-shaped footprints at the bottom surface (\Fig~\ref{fig4}c), and
these footprints consists of the limits of the MFR (i.e., the hook
parts of the QSL footprints) and the footpoints of the field lines
that are undergoing reconnection in the current sheet (mainly the leg
parts), and the latter are believed to correspond to the location of
flare ribbons. Indeed, the shape and evolution of the QSL footprints
are in reasonable consistence with that of the observed flare ribbons
in SDO/AIA 1600~{\AA}, by taking into consideration of the systematic
difference of the photospheric magnetic field between the simulation
and observation. By tracing the movement of the QSL footprints, one
can compute the evolution of the reconnection flux, which is the
magnetic flux as swept by the moving QSL
footprints. \Fig~\ref{recflux_compare} shows that the results, in both
the reconnection flux and the reconnection rate, are comparable to
those derived from evolution of the observed flare
ribbons~\citep{heQuantitativeCharacterizationMagnetic2022}.  Note that
there are also reconnection within the
MFR~\citep{jiangFormationMagneticFlux2021}, and this reconnection
creates field lines with higher twist number than that created by the
initial tether-cutting reconnection. As a result, these field lines
are separated with others by the newly-formed embedded QSLs in each
hook, which are similar to the ones described in an analytic
study~\citep{demoulinThreedimensionalMagneticReconnection1996}.

\subsection{The pre-eruption slow rise phase}
\label{slow_rise}
Although \Fig~\ref{paraevol} shows a typical slow storage of magnetic
free energy to its fast release in eruption, the evolution can
actually be divided into three different stages, that is, a
quasi-static phase, an impulsive rise phase, and a slow rise phase in
between, which is likely to correspond to the observed short period of
coronal loop slow expansion immediately before the eruption. The first
phase from the beginning to around $t=20~t_s$ is truly quasi-static
because, on the one hand, the core field expands with a speed close to
that of the bottom driving speed (\Fig~\ref{paraevol}c), and on the
other hand, at any instant in this phase, if we stop the bottom
driving (i.e., by turning off the rotation of the sunspot), the system
can relax smoothly to an equilibrium with gradual decline of the
kinetic energy. For instance, the kinetic energy keeps decreasing once
the rotation is switched off at $t=20~t_s$. In the quasi-static phase,
almost all the magnetic energy injected from the bottom boundary is
stored in the coronal volume, as shown by the close match of the
energy injection line (the gray curve in \Fig~\ref{paraevol}a) and the
evolution profile of the total magnetic energy. Clearly, the time
scale of the quasi-static evolution depends on that of the bottom
driving. Since we used a surface speed (of about $10$~km~s$^{-1}$)
larger than the actual photospheric motion speed, the quasi-static
phase in our simulation (with a duration of
$20~t_s \approx 0.6$~hours) is much shorter than the realistic one
(nearly $60$ hours). It can be scaled approximately to the realistic
one if we use the realistic speed (for example, on the order of
$0.1$~km~s$^{-1}$) of the photospheric motion, but that demands too
much computing time.

The second phase from around $t=20~t_s$ to the onset time of eruption
(i.e., $28~t_s$) is a slow rise phase, in which the kinetic energy
slowly rises and the magnetic energy evolution begins to deviate
evidently from the magnetic energy injection curve. This phase is also
featured by a relatively large speed of $50\sim 60$~km~s$^{-1}$ as
shown in the expansion rate of the core field. A more important reason
why this phase is different from the first phase in nature is that, if
the bottom driving is turned off at any moment in this phase, the
system will not relax to a static equilibrium (for example, see the
evolution of the kinetic energy when the surface driving is turned off
at $t=23~t_s$ in \Fig~\ref{paraevol}a).  Rather, the kinetic energy
keeps an approximately constant value without decay, and meanwhile the
magnetic energy decreases slowly. Moreover, no matter at which moment
the boundary driving is switched off, the system will always reach an
eruption phase after a short interval of evolution (for instance, see
the case with sunspot rotation stopped at $t=23~t_s$, for which the
magnetic field and current density evolution are shown in
the animation attached for \Fig~\ref{paraevol}). These experiments show that the bottom driving
is not necessary in maintaining the slow rise phase (once it begins),
although it can slightly speed up this process.
The time duration of slow rise phase is around $8~t_s$, which is on
the same order of that of the observed slow rise phase (around
$20$~min).

Why the slow rise phase must proceed for a short interval before the
eruption, regardless of the bottom boundary driving?  This
  is because only when the current sheet reaches the critical
  thickness can the eruption be triggered by reconnection.
In \Fig~\ref{paraevol}b, we also show evolution of the thickness of
the current sheet in the different runs. As can be seen, the current
layer is continually thinned and once its thickness is below the
resolvable limit of the grid resolution (about 3 grid sizes), the
impulsive acceleration of the fast eruption phase begins. When turning
off the bottom driving at $t=23~t_s$, the current layer is still
thinned although the rate is slightly slower, and the eruption is
initiated in the same way when the current sheet reaches the critical
thickness. In contrast, when the boundary driving is stopped at
$t=20~t_s$ before the slow-rise phase begins, the system will relax to
an equilibrium without changing the thickness of the current layer,
and thus cannot develop the current sheet as required for
reconnection.

Although our simulations suggest a slow rise phase analogous to the
observed one, the duration of this phase should be sensitive to the
magnitudes of the resistivity and the bottom boundary driving. If the
resistivity is smaller, the eruption is expected to be delayed since a
thinner current sheet would form and needs more time, thus this phase
would be longer. With a slower driving speed at the bottom
boundary, this phase is also expected to be longer.
In our simulations, both the resistivity and the driving speed are
much larger than the real values, for which it is formidable to
realize at present in numerical simulations.


\section{Conclusions and discussion}\label{con}
In this paper, based on observation and MHD simulations, we have
studied the process of a rotating sunspot in AR~12158 causing a
major solar eruption. Our simulations demonstrated that the continuous
rotation of a major sunspot of the AR leads to an eruption in a way
distinct from the conventional view based on ideal MHD instabilities
of twisted flux rope. It is found that through the successive rotation
of the sunspot the coronal field is sheared with a vertical current
sheet created progressively, and once fast reconnection sets in at the
current sheet, the eruption is instantly triggered, and a highly
twisted flux rope originates from the eruption, forming a CME. 
Therefore, in the two independent simulations, i.e., a data-driven one in {\PaperI} and a data-inspired one here, we found that their eruptions are initiated by essentially the same way: the coronal field is stressed until a central current sheet forms with fast reconnection which triggers and results in eruption. 

The simulation here also revealed explicitly a slow-rise evolution phase
between the quasi-static phase of non-potential magnetic energy
storage and the impulsive acceleration phase of eruption. Such a
slow-rise phase is commonly observed in many eruption events
(including the studied one) for either erupting filaments or coronal
loops but still lacks a physical explanation. Our analysis suggests
that the slow-rise phase is
inherent to the coronal dynamics when close to eruption rather than
controlled by the photospheric driving motions as considered
before~\citep{vrsnakGradualPreeruptivePhase2019}.
Once it begins the slow-rise phase develops for a short time interval (often with a
few tens of minutes), even if the boundary driving is turned off, and
eventually transforms into an eruption. This short phase plays
an important role in building up of the current sheet, since it
accelerates the thinning of the current layer that is built up in the
quasi-static phase until the reconnection sets in.

Why do our simulations support such a scenario rather than that based
on MHD instabilities of pre-eruption MFRs? There are two
reasons. Firstly, the rotation of sunspot is too slow to form a
well-defined MFR in a few days. Observations show that the angular
rotational speed for many sunspots is on average a few degrees per
hour and the total rotation degree is mostly between 40--200$^{\circ}$
over periods of 3--5
days~\citep{brownObservationsRotatingSunspots2003}. Thus, the resulted
magnetic twist by such rotation is far lower than that necessary to
form a well-defined MFR (which needs field lines winding around an
axis with at least one turn, or twist of $360^{\circ}$) and further to
trigger the ideal kink instability (which needs winding number of more
than $1.25$ turns~\citep{hoodKinkInstabilitySolar1979} or twist of
more than $450^{\circ}$). Even though the sunspot in the well known AR~10930 is reported
 to rotate by a very large amount of about
$500^{\circ}$ over five days~\citep{minRotatingSunspotAR2009}, there
were many eruptions in that duration (AR~10930 produced 4 homologous
X-class eruptive flares and a few smaller ones) and thus these
eruptions should release substantially the built-up twist repeatedly,
preventing the continuous formation of a well-defined twisted
rope. Indeed, this is supported by a previous study of
AR~10930~\citep{wangMHDSimulationHomologous2022a}, in which a MHD
simulation was carried out using a similar approach in this paper.  In
that work, the simulation started with a potential magnetic field
reconstructed from the observed magnetogram and then rotational motion
is applied to the positive sunspot of the AR to mimic the observed
rotation. That simulation successfully showed that the sunspot
rotation produced homologous eruptions having reasonable consistency
with observations in relative strength, energy release, spatial
features (such as pre-eruption sigmoid and flare ribbons), and time
intervals of eruptions. In addition, in the simulation the total angle of rotation of the sunspot
until the eruption onset is also very close to the observed
value. The key finding is also similar to this paper: as driven by the
sunspot rotation, current sheet is formed above the main PIL between
the two major magnetic polarities of the AR, and the eruptions are
triggered by fast reconnection in the pre-eruption formed current
sheet.



Secondly, sunspot rotation, although slow, is a very efficient way of
injecting magnetic free energy. For example, by modelling of a solar
flare from 13 May 2005, it has been shown that the sunspot rotation of
the source AR dominates the energy accumulation for the flare event
\citep{kazachenkoSUNSPOTROTATIONFLARE2009}. In fact, such sunspot
rotation alone can store sufficient energy to power a very large
flare. Our simulation demonstrated this point more clearly as shown in
\Fig~\ref{paraevol}a; by a rotation of $90^{\circ}$ (on average for
the whole sunspot), the free energy has been increased by over 50\% of
the potential field energy. As the open field energy is an upper
limiter, which is around $2.5$ times of the potential field energy,
this made it very easy for the field to reach the open field energy,
if the rotation is as large as $300^{\circ}$. The open field is
closely related to the building up of the current sheet, since a
current sheet must be built up before the field becoming fully open,
as consistent with our previous high-resolution simulation based on a
simple bipolar field~\citep{jiangFundamentalMechanismSolar2021}. Thus,
to build up a current sheet needs much less degree of rotation than
that for building up an unstable MFR, as the former is more consistent
with the observation of the rotation degrees.

Nevertheless, there is also a possibility in some events that the
rotation of sunspot drives the formation of the current sheet, but a
confined flare is resulted if the overlying field is strong enough (or
decay slowly enough) to constrain the newly-formed, erupting flux
rope. With continuous rotation, such confined flares might occur
multiple times and jointly build up an ideal unstable MFR that
eventually erupts, as suggested by recent observations
\citep{jamesNewTriggerMechanism2020}. Another possibility is that
during the sunspot rotation, continuous slow reconnection might occur
at near the photosphere, which is often associated with the observed
flux cancellation, can also help to build up unstable MFR before
eruption.
This will need further investigations with more events and simulations.

\begin{acknowledgments}
This work is jointly supported by Shenzhen Science
and Technology Program (Grant No. RCJC20210609104422048), Shenzhen
Technology Project JCYJ20190806142609035, Shenzhen Key Laboratory
Launching Project (No. ZDSYS20210702140800001), Guangdong Basic and
Applied Basic Research Foundation (2023B1515040021) and National
Natural Science Foundation of China (NSFC 42174200). Data from
observations are courtesy of NASA SDO/AIA and the HMI science
teams. The computational work was carried out on TianHe-1(A), National
Supercomputer Center in Tianjin,
China. 
\end{acknowledgments}

\bibliographystyle{aasjournal}
\bibliography{all} 

\end{document}